\documentclass[10pt,final,journal,twocolumn]{IEEEtran}
\usepackage{cite}
\usepackage{amsmath,amssymb,amsfonts}
\usepackage{graphicx}
\usepackage{textcomp}
\usepackage{xcolor}
\usepackage{subcaption}
\def\BibTeX{{\rm B\kern-.05em{\sc i\kern-.025em b}\kern-.08em
    T\kern-.1667em\lower.7ex\hbox{E}\kern-.125emX}}

\usepackage[ruled]{algorithm2e}

\def\A{\boldsymbol{A}}

\def\G{\boldsymbol{G}}
\def\H{\boldsymbol{H}}
\def\I{\boldsymbol{I}}

\def\U{\boldsymbol{U}}
\def\V{\boldsymbol{V}}

\def\Y{\boldsymbol{Y}}
\def\Z{\boldsymbol{Z}}

\def\a{\boldsymbol{a}}

\def\e{\boldsymbol{e}}

\def\g{\boldsymbol{g}}

\def\p{\boldsymbol{p}}

\def\s{\boldsymbol{s}}

\def\u{\boldsymbol{u}}
\def\v{\boldsymbol{v}}

\def\x{\boldsymbol{x}}

\def\bLambda{\boldsymbol{\Lambda}}

\def\btheta{\boldsymbol{\theta}}

\def\bxi{\boldsymbol{\xi}}

\def\argmin#1{\underset{#1}{\textrm{argmin}}}

\def\maxim#1{\underset{#1}{\textrm{max}}}

\def\minim#1{\underset{#1}{\textrm{min}}}

\SetKwInput{kwEvaluate}{Evaluate}
\SetKwInput{kwSort}{Sort}
\SetKwInput{kwInput}{Input}
\SetKwInput{kwOutput}{Output}
\SetKwInput{kwInitialize}{Initialize}
\SetKwInput{kwParameters}{Params.}
\SetKwInput{kwDefaults}{Default init.}
\SetKwRepeat{Do}{do}{while}

\newcommand{\tomt}[1]{\textcolor{black}{#1}}
\newcommand{\tomtb}[1]{\textcolor{black}{#1}}
\newcommand{\tomtr}[1]{\textcolor{black}{#1}}

\begin{document}

\title{
Direction of Arrival Estimation and Phase-Correction for Non-Coherent Sub-Arrays:
\\A Convex Optimization Approach
}

\author{
    Tom Tirer and Oded Bialer 
    \thanks{
    Tom Tirer is with the Center for Data Science, New York University, New York, USA. This work was done when he was with General Motors Advanced Technical Center, Herzliya 46733, Israel (e-mail: tirer.tom@gmail.com).
    
    Oded Bialer is with General Motors Advanced Technical Center, Herzliya 46733, Israel (e-mail: oded.bialer8@gmail.com).
    }
}

\maketitle

\begin{abstract}
\tomtb{Estimating the direction of arrival (DOA) of sources is an important problem in \tomtb{aerospace and} vehicular 
communication, localization and radar. In this paper, we}  
consider a challenging multi-source 
DOA 
estimation task, 
where the receiving antenna array is composed of non-coherent sub-arrays, i.e., sub-arrays that observe different unknown phase shifts at every snapshot 
\tomtb{(e.g., due to waiving the demanding synchronization of local oscillators across the entire array).} 
We formulate this problem as the reconstruction of 
joint sparse and low-rank matrices, 
and solve 
the problem's 
convex relaxation.  
To 
scale the optimization complexity with the number of snapshots 
better than general-purpose solvers, 
we design an optimization scheme, based on integrating the alternating direction method of multipliers and the accelerated proximal gradient techniques, that exploits the structure of the problem. 
While the DOAs can be estimated from the solution of the aforementioned convex problem, we further show how an improvement is obtained if, instead, one estimates from this solution only the sub-arrays' phase shifts.
This is done using another, computationally-light, convex relaxation that is practically tight. Using the estimated phase shifts, ``phase-corrected" observations are created and a final plain (``coherent") DOA estimation method can be applied. Numerical experiments show 
the performance advantages
of the proposed strategies over existing methods. 
\end{abstract}



\section{Introduction}\label{sec:INTRO}

Estimating the directions of arrival (DOAs) of radio frequency sources using an array of antenna elements 
is 
a popular problem 
that 
attracts 
much interest in different disciplines \cite{krim1996two}, 
\tomtb{and is important for various \tomtb{aerospace and}  vehicular applications, such as 
communication, localization, and radar \cite{li2001moving,xu2017optimal,arenas2019direction,del2019network,fascista2016localization,
wang2019assistant,bilik2016automotive}.} 
When the elements of the receiving antenna array are {\em coherent},\footnote{Coherent elements are 
antenna elements that are synchronized in phase and frequency to the same local oscillator. 
Therefore, ignoring the noise, the phase differences between their observations are only due to the difference in the propagation delays between the location of the sources and the location of the elements.}
the DOA estimation performance improves
\tomt{(in both accuracy and resolution)}
with the increase of the array aperture 
\cite{van2004optimum, stoica1988music}.
However, maintaining the coherence of all the elements 
of a large array 
is very demanding, especially at high carrier frequency, as it requires expensive hardware and a complicated calibration process that is prone to errors.

A possible alternative, with lower hardware complexity and cost, 
is to split a large array into several smaller sub-arrays, such that the coherence of the elements is kept only within each sub-array (using low-cost unsynchronized local oscillators 
for the different sub-arrays)
while the large aperture of the entire array is exploited using signal processing techniques \cite{tirer2020method}. 
\tomtb{This design approach has been used recently {\em in practice} for obtaining a wide aperture automotive MIMO radar with higher resolution than existing state-of-the-art alternatives \cite{bialer2021super}. 
However, the system presented in \cite{bialer2021super} still has limited exploitation of the full antenna array for DOA estimation.}  

A common approach 
for estimating the DOAs from non-coherent sub-arrays 
is based on {\em non-coherent processing} of the sub-arrays, i.e., handling the sub-arrays as if they observe different signals \cite{rieken2004generalizing, wen2014improved, suleiman2018non}. Yet, while non-coherent processing bypasses the need to estimate the sub-arrays' phase shifts, it does not use the phase information between the sub-arrays, and hence does not exploit the potential aperture of the entire array.

\tomtr{
Alternatively, 
there exist approaches, typically 
designed for the self-calibration task, 
that estimate 
gain-phase offsets between antenna elements or sub-arrays jointly
with the DOAs. 
Some of these works 
are based on subspace methods (such as variants of MUSIC \cite{schmidt1986multiple}) applied on the sample covariance of the entire array, with modifications that allow to get rid of the nuisance 
offsets parameters \cite{pesavento2001direction, see2004direction, parvazi2011direction, friedlander1991direction, swindlehurst2001exploiting}.
However, to identify the ``signal space", these methods require a large number of snapshots (temporal observations) {\em with the same phase offsets} between the sub-arrays, while
in the problem considered in this paper there are
{\em different phase offsets in each snapshot} 
(e.g., due to unsynchronized local oscillators).
A few other self-calibration techniques \cite{liu2013unified,ling2015self,hung2017low}, which also assume that the sub-arrays' offsets remain constant at all the snapshots, are based on sparsity rather than on subspace decomposition. 
As such, they may be adapted to the problem that is considered in this paper, though, only for the single snapshot case.
Yet, even in this case 
we empirically show that the methods that we develop 
here 
significantly outperform such an adaptation.} 

\tomt{
The variability of the phase shifts across snapshots hints that solutions to the considered 
DOA estimation 
problem may be inspired by 
approaches that are applicable to the single snapshot case (unlike subspace methods). 
Such methods include minimization 
of the negative log-likelihood function \cite{ziskind1988maximum, bresler1986exact, feder1988parameter}, 
sparse signal reconstructions \cite{jeffs1998sparse, malioutov2005sparse,yang2018sparse}, 
and machine learning techniques \cite{bialer2019performance}.
Indeed, 
to advance the performance of DOA estimation for non-coherent sub-arrays, 
the 
recent 
work in \cite{tirer2020method,tirer2020effective} has 
proposed an approximate maximum likelihood approach, 
under the assumption that the number of sources is known. 
However, 
this method  
is based on non-convex optimization with respect to an optimization variable whose dimension is the number of sources. Therefore, it requires a 
heavy (practically intractable) 
multidimensional search, or, when solved using iterative schemes (such as \cite{ziskind1988maximum, bresler1986exact, feder1988parameter}), it 
depends on the quality of the initialization.}

In this paper, we propose a novel approach for estimating the DOAs from either a single snapshot or multiple snapshots of non-coherent sub-arrays.%
\footnote{\tomtb{A few preliminary single-snapshot results appeared in our conference version \cite{tirer2020direction}. The current extended version generalizes the problem to multiple snapshots, designs an efficient optimization algorithm (which is indispensable for handling cases with multiple snapshots), replaces the naive phase shift estimation in \cite{tirer2020direction} with a well-justified sub-problem that admits a sophisticated tight convex relaxation, and includes a more extensive empirical study.}} 
Our approach 
is based on convex optimization, and thus does not rely on good initializations and can easily handle scenarios with a large (possibly unknown) number of sources. 
Borrowing ideas from sparsity-based DOA estimation for coherent arrays \cite{malioutov2005sparse} and from low-rank-based blind deconvolution \cite{ahmed2013blind}, we formulate the problem as the reconstruction of joint sparse and low-rank matrices and solve the problem's convex relaxation.
The complexity of solving the formulated problem with general-purpose solvers \cite{grant2011cvx,toh1999sdpt3} grows extremely with the number of snapshots. Therefore, we derived a computationally efficient optimization algorithm that is based on combining the alternating direction method of multipliers (ADMM) \cite{boyd2011distributed} and the accelerated proximal gradient (FISTA) \cite{beck2009fast} techniques. 

\tomt{
Given the solution of the convex problem, the DOAs can be directly estimated.
Yet, we present another strategy that further exploits the model's structure and improves the results:
instead of estimating the DOAs, the aforementioned solution is used to estimate 
{\em only the nuisance sub-arrays' phase shifts}.}  
This is done using another, computationally-light, convex relaxation that is practically tight. With these estimates, we 
create ``phase-corrected" observations and can use a final plain (``coherent") DOA estimation method, such as MUSIC \cite{schmidt1986multiple} or $\ell_1$-norm-based sparsity method \cite{malioutov2005sparse}. 
To the best of our knowledge, no similar 
two-stage procedure has been proposed and examined in the 
literature so far. 
We examine the proposed strategies using various numerical experiments, which show that they 
\tomt{
outperform non-coherent processing methods, other sparsity and low-rank based techniques,
and also the method in  \cite{tirer2020method}.}


\section{Problem Formulation}
\label{sec:FORM}

Consider $Q$ far-field sources, located at unknown angles $\btheta \triangleq [\theta_1, \ldots, \theta_Q]^T$ and transmitting unknown narrowband signals that impinge on an $M$-element receiving array.  
Let the array be composed of $L$ sub-arrays, where the $\ell$th sub-array consists of $M_\ell$ elements 
($\sum \limits_{\ell=1}^{L} M_\ell = M$).
We assume that each sub-array is coherent, while different sub-arrays are non-coherent, i.e., 
there is an unknown phase shift between sub-arrays. 
We assume that $N$ snapshots are observed 
and that there are different unknown phase shifts at every snapshot.

The waveform observed 
at the $n$th snapshot of the $\ell$th sub-array 
can be formulated as
\begin{align}
\label{Eq_subarray_model}
&\x_{\ell}(n) = \mathrm{e}^{-j \phi_{\ell}(n)} \tilde{\A}_{\ell}(\btheta)\tilde{\s}(n) + \e_{\ell}(n), \\ \nonumber
&\hspace{25mm} \ell=1,\ldots, L;  \,\, n=1,\ldots, N,
\end{align}
where $\phi_{\ell}(n)$ is the unknown phase in the $\ell$th sub-array at the $n$th snapshot, 
$\tilde{\A}_{\ell}(\btheta) \triangleq [\a_{\ell}(\theta_1), \ldots, \a_{\ell}(\theta_Q)] \in \mathbb{C}^{M_\ell \times Q}$ 
is the $\ell$th sub-array manifold, where each column $\a_{\ell}(\theta)$ is the $\ell$th sub-array response to a signal which arrives at angle $\theta$. 
The vector 
$\tilde{\s}(n) \in \mathbb{C}^{Q \times 1}$ 
represents the unknown signals at the $n$th snapshot, and 
$\e_{\ell}(n) \in \mathbb{C}^{M_\ell \times 1}$ 
represents white, zero-mean, circular complex Gaussian noise. 
To simplify the formulation, we assume that the noise variance $\sigma^2$ is known, and equal at all receivers (the extension to
non equal variances is straightforward).

Regarding the steering vector $\a_{\ell}(\theta)$, 
assuming planar geometry where the axis origin (i.e., the reference point) is defined in the middle of the entire array, and that the angle of the (far-field) source, $\theta$, is measured with respect to the array boresight, then 
\begin{align}
\label{Eq_subarray_response}
[\a_{\ell}(\theta)]_i = \eta_{\ell,i}(\theta) \mathrm{e}^{j \frac{2\pi}{\lambda_f}(x_{\ell,i}\mathrm{sin}\theta + y_{\ell,i}\mathrm{cos}\theta)},
\end{align}
where $\lambda_f$ is the signal wavelength, $(x_{\ell,i},y_{\ell,i})$ denotes the antenna location, and $\eta_{\ell,i}$ denotes the antenna element pattern, which is assumed to be known. 
For example, for linear arrays of omnidirectional sensors we have that 
$\eta_{\ell,i}(\theta)=1$ and $y_{\ell,i}=0$.

To conclude this section, 
the problem at hand is to estimate the DOAs $\btheta$ from the sub-arrays' observations $\{ \x_{\ell}(n) \}$ given in \eqref{Eq_subarray_model}, where $\{\phi_{\ell}(n)\}$ and $\{ \tilde{\s}(n) \}$ are nuisance parameters.


\section{The Proposed Method}
\label{Sec:METHOD}

We propose to tackle the problem from the sparse signal reconstruction  perspective \cite{malioutov2005sparse}. Namely, we define a DOA grid of length $N_\theta \gg Q$ (containing different hypotheses of a  scalar DOA) and for each sub-array create the 
measurement matrix  $\A_{\ell} \triangleq [\a_{\ell}(\theta_1), \ldots, \a_{\ell}(\theta_{N_\theta})] \in \mathbb{C}^{M_\ell \times N_\theta}$.
The observation model in \eqref{Eq_subarray_model} can now be reformulated as
\begin{align}
\label{Eq_subarray_model_sparse}
&\x_{\ell}(n) = p_{\ell}^*(n) \A_{\ell} \s(n) + \e_{\ell}(n), \\ \nonumber
&\hspace{25mm} \ell=1,\ldots, L;  \,\, n=1,\ldots, N,
\end{align}
where $p_{\ell}(n) \triangleq \mathrm{e}^{j \phi_{\ell}(n)}$ (the superscript $^*$ denotes the complex conjugate), and 
$\s(n) \in \mathbb{C}^{N_\theta \times 1}$ 
obeys $\|\s(n)\|_0=Q$, where $\|\cdot\|_0$ is the $\ell_0$ pseudo-norm that counts the number of non-zero entries of the vector.
Notice that all the vectors $\{\s(n)\}$ have the same support, i.e., the $Q$ non-zero values are located at the same entries at each of them (because we assume that the DOAs are the same in all the snapshots). 
Moreover, observe that the complicated non-linear model in \eqref{Eq_subarray_model} is replaced with the model in \eqref{Eq_subarray_model_sparse} that is bilinear with respect to the unknown parameters $\p(n) \triangleq [p_{1}(n), \ldots, p_{L}(n)]^T$ and $\s(n)$. This suggests that, similarly to the ``lifting" strategy used in \cite{ahmed2013blind} for the blind deconvolution problem (which is also bilinear), we can express \eqref{Eq_subarray_model_sparse} as a linear model with respect to a rank-1 matrix $\Z_n \triangleq \s(n) \p^H(n)  \in \mathbb{C}^{N_\theta \times L}$. Indeed, 
\begin{align}
\label{Eq_subarray_model_sparse_rank1}
&\x_{\ell}(n) = \A_{\ell} \Z_n[:,\ell] + \e_{\ell}(n), \\ \nonumber
&\hspace{25mm} \ell=1,\ldots, L;  \,\, n=1,\ldots, N,
\end{align}
where $\Z_n[:,\ell]$ denotes the $\ell$th column of $\Z_n$.
Note that $\Z_n$ is not only rank-1, but also ``row-sparse", i.e., only $Q$ of its rows are non-zero. 
Importantly, since $\{ \s(n) \}$ have the same sparsity support, we have that $\{ \Z_n \}$ have the same ``row-sparse" pattern. 
Note also that, in practice, the number of sources, $Q$, may not be known in advance.

\tomtr{Motivated by the theoretical work on the estimation benefits that come from tight signal modeling \cite{chandrasekaran2012convex}, we would like to use both the sparsity and the low-rankness properties of the true $\{ \Z_n \}$.} 
Therefore, a natural non-convex optimization problem that arises from 
the observation model in \eqref{Eq_subarray_model_sparse_rank1} and the row-sparsity and low-rankness priors is 
\begin{align}
\label{Eq_opt_prob_nonconvex}
&\minim{\{\Z_n\} \in \mathbb{C}^{N_\theta \times L}} \,\,\,  \beta \| [ \Z_1, \ldots, \Z_N ] \|_{0,2} + \mu \sum_{n=1}^N  \mathrm{rank}(\Z_n) \nonumber\\
&\hspace{7mm} \mathrm{s.t.} \,\,\,\, \sum_{n=1}^{N} \sum_{\ell=1}^{L} \left \|\x_{\ell}(n) - \A_{\ell} \Z_n[:,\ell] \right \|_2^2 \leq C N M \sigma^2,
\end{align}
where $\beta$, $\mu$ and $C$ are positive hyper-parameters, $\|\cdot\|_2$ stands for the Euclidean norm, and $\|\cdot\|_{0,2}$ is a pseudo-norm that counts the number of non-zero rows of the matrix (or equivalently, counts the number of rows whose Euclidean norm is non-zero).
The hyper-parameters $\beta$ and $\mu$ balance between the row-sparsity and the low-rankness terms. Clearly, it is enough to fix one of them to 1 and tune only the other. Yet, we define both of them to simplify the presentation (e.g., in the experiments section, where we will compare our approach to methods that use only the row-sparsity or the low-rankness).
As for the hyper-parameter $C$, note that its tuning is quite interpretable: for sufficiently large (yet still moderate) $C$, e.g., $C=2$, the feasible set includes the true unknown $\{\Z_n\}$ with high probability. 
Note that we currently ignore the fact that $\{p_\ell(n)\}$ have unit magnitude. 
This is due to the difficulty in expressing such constraints in terms of $\{ \Z_n \}$ (each $\Z_n = \s(n) \p^H(n)$ depends also on the unknown $\s(n)$).  
Yet, this information will be used later on.


Hypothetically, 
given a solution to \eqref{Eq_opt_prob_nonconvex}, one can 
construct from it a 
``spectrum" (score for each of the $N_\theta$ DOA hypotheses), and estimate the DOAs as the angles associated with indices of the peaks of the magnitude of the spectrum. 
However, \eqref{Eq_opt_prob_nonconvex} is a non-convex problem (due to its objective) that is not tractable to solve.   
Therefore, 
to obtain a convex problem that can be efficiently solved we perform convex relaxation on \eqref{Eq_opt_prob_nonconvex}. Following the common practice \cite{chen1994basis,chen2001atomic,fazel2003matrix}, we replace $\mathrm{rank}(\Z_n)$ with the nuclear norm $\|\Z_n\|_{*}$ that sums the singular values of the matrix, and replace $\ell_0$-type pseudo-norms with their $\ell_1$-norm counterparts, i.e., we replace $\| [ \Z_1, \ldots, \Z_N ] \|_{0,2}$ with $\| [ \Z_1, \ldots, \Z_N ] \|_{1,2} \triangleq \sum \limits_{i=1}^{N_\theta} \sqrt{ \sum_{n=1}^{N} \sum_{\ell=1}^{L} |Z_n[i,\ell]|^2 }$. 
Therefore, we get the following convex optimization problem
\begin{align}
\label{Eq_opt_prob_convex}
&\minim{\{\Z_n\} \in \mathbb{C}^{N_\theta \times L}} \,\,\,  \beta \| [ \Z_1, \ldots, \Z_N ] \|_{1,2} + \mu \sum_{n=1}^N  \| \Z_n \|_* \nonumber\\
&\hspace{7mm} \mathrm{s.t.} \,\,\,\, \sum_{n=1}^{N} \sum_{\ell=1}^{L} \left \|\x_{\ell}(n) - \A_{\ell} \Z_n[:,\ell] \right \|_2^2 \leq C N M \sigma^2.
\end{align}

The above convex program 
can be directly solved by the CVX Matlab package \cite{grant2011cvx}, which applies the SDPT3 solver \cite{toh1999sdpt3}, whose computational complexity is polynomial in $N_\theta$, $L$ and $N$. 
Since typically $L$ (the number of sub-arrays) is small, we observed that this procedure has tractable runtime for a single snapshot $N=1$ (it took 
several seconds in our experiments with Intel-i7 laptop).
However, for multiple snapshots the runtime significantly grows.
To resolve this issue, in Section~\ref{sec:admm} we design a first-order optimization scheme for obtaining $\{ \Z_n \}$ (i.e., a scheme that is based only on first-order derivatives, unlike \cite{toh1999sdpt3}), which exploits the structure of the problem.
Namely, it 
is based on integrating the ADMM \cite{boyd2011distributed} and FISTA \cite{beck2009fast} techniques 
and utilizing 
the fact that both $\| \cdot \|_{1,2}$ and $\| \cdot \|_*$ have proximal mappings with analytical solutions.

Following, 
in Sections~\ref{sec:Prop1} and \ref{sec:Prop2},
we present two strategies for estimating the DOAs, based on the solution of \eqref{Eq_opt_prob_convex}, denoted by $\{ \hat{\Z}_n \}$. 

\subsection{First Strategy: Direct DOA Estimation from $\{ \hat{\Z}_n \}$}
\label{sec:Prop1}

We start by noticing that
although the nuclear norm promotes low-rankness of each $\hat{\Z}_n$, it does not strictly impose rank-1 (recall that the true $\Z_n$ equals $\s(n)\p^H(n)$). To strictly impose it, 
we apply 
the singular value decomposition (SVD) on each $\hat{\Z}_n$.
Namely, we decompose $\hat{\Z}_n =\U_n \bLambda_n \V_n^H$, where $\U_n \in \mathbb{C}^{N_\theta \times N_\theta}$ and $\V_n \in \mathbb{C}^{L \times L}$ are unitary matrices and $\bLambda_n$ is a rectangular diagonal matrix.
Now, the rank-1 approximation of $\hat{\Z}_n$ is given by $\hat{\Z}_n^{\mathrm{R1}}=\Lambda_n[1,1]{\U_n[:,1]}\V_n[:,1]^H$. 
Next, similar to the common approach in methods that are based on group-sparsity \cite{malioutov2005sparse,yang2018sparse}, a spectrum, $\bxi \in \mathbb{R}^{N_\theta}$, is obtained by applying row-wise $\ell_2$-norm on $[ \hat{\Z}_1^{\mathrm{R1}}, \ldots, \hat{\Z}_N^{\mathrm{R1}} ]$. Namely, 
\begin{align}
\label{Eq_prop1}
\xi[i]=\sqrt{ \sum_{n=1}^{N} \sum_{\ell=1}^{L} \left | \hat{Z}_n^{\mathrm{R1}}[i,\ell] \right | ^2 } 
\end{align}
is the score for the DOA hypothesis $\theta_i$ ($i=1,\ldots,N_\theta$). 
The DOAs are estimated as the peaks (highest local maxima) of $\bxi$. 
\tomtr{In the experiments section we demonstrate that the rank-1 approximation step that is described here is indeed beneficial.}

\subsection{Second Strategy: Phase-Correction + ``Coherent" DOA Estimation}
\label{sec:Prop2}

Until now we 
did not use 
the fact that the true $\{p_\ell(n)\}$ have unit magnitude 
(since 
$p_{\ell}(n) = \mathrm{e}^{j \phi_{\ell}(n)}$) 
and only their phases are unknown. 
To utilize this information, we propose 
a new {\em two-stage} 
approach, where in the first stage we estimate from $\{ \hat{\Z}_n \}$ only the sub-arrays' phase shifts 
and in the second stage we estimate the DOAs 
based on the estimated phase shifts.
This strategy, further exploits the model's structure and improves the results in our experiments compared to the strategy in Section~\ref{sec:Prop1}.

In the first stage, 
only the phase shifts $\{\phi_{\ell}(n)\}_{\ell=1}^L$ 
are estimated from 
each $\hat{\Z}_n$.
Recalling that the true $\Z_n$ obeys $\Z_n = \s(n)\p^H(n) \in \mathbb{C}^{N_\theta \times L}$, note that $\Z_n^H\Z_n = \p(n) \s^H(n) \s(n) \p^H(n) = \|\s(n)\|_2^2 \p(n) \p^H(n)$.
Using 
the Cauchy-Schwarz inequality, we have
\begin{align}
\label{Eq_opt_optimal_alpha}
\v^H \Z_n^H\Z_n \v &= \|\s(n)\|_2^2 | \p^H(n) \v |^2  \nonumber\\
& \leq \|\s(n)\|_2^2 \| \p(n) \|_2^2 \| \v \|_2^2. 
\end{align}
Clearly, the upper bound is met for $\v=\p(n)$, or multiplications of $\p(n)$ by a scalar, which do not change the phase difference between the entries. 
Thus, if the true $\Z_n$ was given, then the phases of $\p(n)$, i.e., $\{\phi_{\ell}(n)\}_{\ell=1}^L$, could have been obtained from the phases of 
the dominant eigenvector of $\Z_n^H\Z_n$. 
In practice, the true $\Z_n$ is unknown, so $\hat{\Z}_n$ may be used instead.
However, since $\hat{\Z}_n$ is not identical to the true $\Z_n = \s(n)\p^H(n)$, rather than computing the (plain) dominant eigenvector of $\hat{\Z}_n^H \hat{\Z}_n$, we should 
look for $\v$ that maximizes $\v^H \hat{\Z}_n^H \hat{\Z}_n \v$ while 
{\em explicitly} enforcing the unit 
magnitude constraints $|v[\ell]|=1$ (that hold for $\v=\p(n)$).

Following the above arguments, we propose to estimate the phases $\{\phi_{\ell}(n)\}_{\ell=1}^L$ from $\hat{\Z}_n$ using
the following optimization problem, which is non-convex due to the nonlinear equality constraints
\begin{align}
\label{Eq_opt_prob_nonconvex_alpha}
&\maxim{ \v \in \mathbb{C}^{L \times 1}} \,\,\,  \v^H  \hat{\Z}_n^H \hat{\Z}_n \v \nonumber\\
&\hspace{5mm} \mathrm{s.t.} \,\,\,\, |v[\ell]|=1, \,\, \ell=1,\ldots,L.
\end{align}
Again, the intuition behind \eqref{Eq_opt_prob_nonconvex_alpha} is that if $\hat{\Z}_n$ is close to $\Z_n = \s(n)\p^H(n)$, then the maximizer of \eqref{Eq_opt_prob_nonconvex_alpha} is expected to be close to the true $\p(n)$ (up to some constant phase that can be absorbed in $\s(n)$ and hence will not affect the DOA estimation). 
Now, observe that \eqref{Eq_opt_prob_nonconvex_alpha} is equivalent to
\begin{align}
\label{Eq_opt_prob_nonconvex_alpha2}
&\maxim{ \v \in \mathbb{C}^{L \times 1}} \,\,\,  \mathrm{Tr}( \hat{\Z}_n^H \hat{\Z}_n \v \v^H ) \nonumber\\
&\hspace{5mm} \mathrm{s.t.} \,\,\,\, v^*[\ell]v[\ell]=1, \,\, \ell=1,\ldots,L, 
\end{align}
which motivates using a semidefinite programming (SDP) relaxation approach in order to mitigate the quadratic equalities \cite{luo2010semidefinite}. First, the problem is equivalently stated using the optimization variable $\V \in \mathbb{C}^{L \times L}$ (restricted to be a positive semidefinite matrix of rank-1, like $\v\v^H$)
\begin{align}
\label{Eq_opt_prob_nonconvex_alpha3}
&\maxim{ \V \in \mathbb{C}^{L \times L}} \,\,\,  \mathrm{Tr}( \hat{\Z}_n^H \hat{\Z}_n \V ) \nonumber\\
&\hspace{5mm} \mathrm{s.t.} \,\,\,\, \mathrm{diag}(\V)=\mathbf{1}_L, \,\, \mathrm{rank}(\V)=1, \,\, \V \succcurlyeq 0.
\end{align}
Here, $\mathrm{diag}(\V)$ denotes the elements on the diagonal of $\V$, $\mathbf{1}_L$ is an $L \times 1$ vector of ones, and $\V \succcurlyeq 0$ restricts the optimization to the positive semidefinite cone. 
Next, SDP relaxation is applied simply by dropping the rank constraint, which yields the following convex program 
\begin{align}
\label{Eq_opt_prob_convex_alpha}
&\maxim{ \V \in \mathbb{C}^{L \times L}} \,\,\,  \mathrm{Tr}( \hat{\Z}_n^H \hat{\Z}_n \V ) \nonumber\\
&\hspace{5mm} \mathrm{s.t.} \,\,\,\, \mathrm{diag}(\V)=\mathbf{1}_L, \,\, \V \succcurlyeq 0.
\end{align}
An interesting side-note is that, unlike the convex relaxation of \eqref{Eq_opt_prob_nonconvex} to \eqref{Eq_opt_prob_convex}, where the rank operation is replaced with the nuclear norm, in \eqref{Eq_opt_prob_convex_alpha} there is no point in using the nuclear norm of $\V$, because $\mathrm{diag}(\V)=\mathbf{1}_L$ and $\V \succcurlyeq 0$ imply fixed $\|\V\|_*=\mathrm{Tr}(\V)=L$ for any feasible $\V$.

The 
convex program \eqref{Eq_opt_prob_convex_alpha} can be directly solved by the CVX Matlab package \cite{grant2011cvx} and is 
quite
fast for typically small $L$ (it took 
about 0.2s
in our experiments with Intel-i7 laptop).
\tomtr{Yet, as this problem is solved per snapshot, 
the overall complexity increases with the number of snapshots. The runtime of solving \eqref{Eq_opt_prob_convex_alpha} can be drastically reduced by a 
dedicated ADMM scheme that we develop in the appendix, and will be referred to in this paper as Algorithm~\ref{opt_alg_sync} (as presented in the appendix).
In our experiments, Algorithm~\ref{opt_alg_sync} solved \eqref{Eq_opt_prob_convex_alpha} in about 0.02s (10 times faster than CVX).}

Let us denote the solution to \eqref{Eq_opt_prob_convex_alpha} by $\hat{\V}_n$. In general, the solution of an SDP relaxation may not be of rank 1 \cite{luo2010semidefinite}. In this case, an additional rank-1 approximation is required for obtaining $\hat{\V}_n \approx \hat{\v}_n \hat{\v}_n^H$ (e.g., using SVD or the power method). Though, then we may not have that $|\hat{v}_n[\ell]|=1$. Interestingly, {\em in all our experiments we observed that the solutions $\{ \hat{\V}_n \}$ were rank-1 matrices} (the second large eigenvalue of $\hat{\V}_n$ was at least 1e-6 times smaller than the maximal eigenvalue).
In other words, we observed that the SDP relaxation is practically tight and provides an optimal solution to the original non-convex problem \eqref{Eq_opt_prob_nonconvex_alpha}.

The above outstanding phenomenon relates to the theory in \cite{bandeira2017tightness} (and its follow-up \cite{zhong2018near}). 
This paper considers a problem that is similar to \eqref{Eq_opt_prob_nonconvex_alpha}. Instead of $\v^H \hat{\Z}_n^H \hat{\Z}_n \v$, the objective there is $\v^H \H \v$, where $\H$ equals $\p \p^H$ (with $|p[\ell]|=1$) plus a Hermitian complex Gaussian noise matrix. Tight SDP relaxation is proved in \cite{bandeira2017tightness} if the noise level is small enough. 
In our case, the role of the Gaussian noise matrix is taken by the mismatch between $\hat{\Z}_n^H \hat{\Z}_n$ and the true $\Z_n^H \Z_n$. Yet, experiments show that our SDP relaxation is still tight, even at low signal to noise ratio.

The outcome of the first stage is the estimated phase shifts, $\{\hat{\phi}_{\ell}(n)\}$, obtained from $\{\hat{\v}_n\}$ 
by 
\begin{align}
\label{Eq_phase_est}
\hat{\phi}_{\ell}(n) = \angle \hat{v}_n[\ell], \,\,\,\, \ell=1,\ldots, L;  \,\, n=1,\ldots, N,
\end{align}
where $\angle x$ stands for the phase of the complex number $x$. 

In the second stage of the proposed strategy, we estimate the DOAs as follows. 
Looking back on
the model in \eqref{Eq_subarray_model}, we use the estimated phase shifts to obtain the ``phase-corrected" observations of the entire array, denoted by $\{\hat{\overline{\x}}(n)\}$, that can be (approximately) modeled as
\begin{align}
\label{Eq_array_model_phase_corrected}
&\hat{\overline{\x}}(n) \triangleq 
\begin{bmatrix}
\mathrm{e}^{j \hat{\phi}_{1}(n)}\x_{1}(n) \\ \vdots \\ \mathrm{e}^{j \hat{\phi}_{L}(n)}\x_{L}(n) 
\end{bmatrix}
\approx \tilde{\A}(\btheta) \tilde{\s}(n) + \overline{\e}(n), \\ \nonumber
&\hspace{45mm} n=1,\ldots, N,
\end{align}
where $\tilde{\A}(\btheta) \triangleq \left [ \tilde{\A}_1^T(\btheta) \ldots \tilde{\A}_L^T(\btheta) \right ]^T \in \mathbb{C}^{M \times Q}$ 
and $\overline{\e}(n) \in \mathbb{C}^M$ 
represents white, zero-mean, circular  complex Gaussian noise, as the noise $\{ \e_\ell(n) \}$ in \eqref{Eq_subarray_model} (because multiplying $\e_\ell(n)$ by $\mathrm{e}^{j \hat{\phi}_{\ell}(n)}$ does not change its statistics). 
As mentioned in previous sections, 
note that in each specific snapshot $n$, only the differences between the phase shifts estimates $\{\hat{\phi}_{\ell}(n)\}_{\ell=1}^L$ matter, as 
the mean (across $\ell$) of the estimated phases can be absorbed 
into the unknown vector $\tilde{\s}(n)$.

At this point, 
plain DOA estimation methods for {\em coherent array} 
can be applied on the ``phase-corrected" snapshots $\{ \hat{\overline{\x}}(n) \}$, modeled in \eqref{Eq_array_model_phase_corrected}. 
In our experiments we use the celebrated MUSIC method \cite{schmidt1986multiple} when $N$, the number of snapshots, allows it (i.e., for $N>Q$, where $Q$ is the number of sources). 
In the experiments that consider the single snapshot case ($N=1$), MUSIC cannot be applied. Instead, we use the popular $\ell_1$-norm sparsity method from \cite{malioutov2005sparse}.

For completeness, let us briefly describe this sparsity-based technique for $N=1$. 
Using the same notations of $\s$ and $\A_\ell$ that are used in \eqref{Eq_subarray_model_sparse} (without the snapshot index $n$ that is unnecessary now), we reformulate \eqref{Eq_array_model_phase_corrected} as
\begin{align}
\label{Eq_array_model_phase_corrected_sparse}
\hat{\overline{\x}} \approx \A \s + \overline{\e},
\end{align}
where $\A \triangleq \left [ \A_1^T \ldots \A_L^T \right ]^T  \in \mathbb{C}^{M \times N_\theta}$, 
and estimate $\s$ as the minimizer of the following convex program
\begin{align}
\label{Eq_opt_prob_convex_phase_corrected}
&\minim{\s} \,\,\, \|\s\|_{1} \nonumber\\
&\mathrm{s.t.} \,\,\,\, \left \| \hat{\overline{\x}} - \A \s \right \|_2^2 \leq C M \sigma^2,
\end{align}
where $C$ is a positive hyper-parameter and $\|\cdot\|_1$ is the $\ell_1$-norm.
This problem can be solved 
relatively fast 
using CVX \cite{grant2011cvx}, and we denote by $\hat{\s}_{\mathrm{cvx}}$ this solution. 
Finally, as done in \cite{malioutov2005sparse}, we estimate the DOAs as the peaks of the magnitude of the $N_\theta \times 1$ vector $\hat{\s}_{\mathrm{cvx}}$.


\subsection{First-Order Optimization Scheme for Obtaining $\{ \hat{\Z}_n \}$}
\label{sec:admm}

In this section we propose an optimization scheme for obtaining $\{ \hat{\Z}_n \}$, which is significantly faster than applying general-purpose solvers for \eqref{Eq_opt_prob_convex}.
In essence, the designed scheme exploits the fact that both $\| \cdot \|_{1,2}$ and $\| \cdot \|_*$ have proximal mappings with analytical solutions (see, e.g., the overview in \cite{parikh2014proximal}). Namely, for a given $\tilde{\Z} \in \mathbb{C}^{N_\theta \times NL}$, the solution of 
\begin{align}
\label{Eq_opt_prox_ell12}
\tilde{\G}=\argmin{\G \in \mathbb{C}^{N_\theta \times NL}} \,\,\,  \tilde{\beta} \| \G \|_{1,2} + \| \G - \tilde{\Z} \|_F^2,
\end{align}
where $\|\cdot\|_F$ denotes the Frobenius norm, is given by the following soft-thresholding 
\begin{align}
\label{Eq_opt_prox_ell12_solution}
&\tilde{\G}[i,:] = \begin{cases}
\tilde{\Z}[i,:] - \frac{\tilde{\beta}}{2} \frac{\tilde{\Z}[i,:]}{\| \tilde{\Z}[i,:] \|_2}, \,\,\, &\| \tilde{\Z}[i,:] \|_2 > \frac{\tilde{\beta}}{2} \\
[0,\ldots, 0], \,\,\, &\| \tilde{\Z}[i,:] \|_2 \leq \frac{\tilde{\beta}}{2}
\end{cases}, \\ \nonumber
&\hspace{5mm} i=1,\ldots, N_\theta.
\end{align}
Similarly, for a given $\tilde{\G}_n \in \mathbb{C}^{N_\theta \times L}$, the solution of 
\begin{align}
\label{Eq_opt_prox_nuc}
\tilde{\Z}_n =\argmin{\Z_n \in \mathbb{C}^{N_\theta \times L}} \,\,\,  \tilde{\mu} \| \Z_n \|_{*} + \| \Z_n - \tilde{\G}_n \|_F^2
\end{align}
is obtain by soft-thresholding the singular values of $\tilde{\G}_n$ by $\tilde{\mu}/2$. Specifically, compute the SVD 
\begin{align}
\label{Eq_opt_soft_thr1}
\tilde{\G}_n=\U_n \bLambda_n \V_n^H,
\end{align}
and soft-threshold the diagonal of $\bLambda_n$
\begin{align}
\label{Eq_opt_soft_thr2}
\tilde{\Lambda}_n[\ell,\ell] = \begin{cases}
{\Lambda}_n[\ell,\ell] - \frac{\tilde{\mu}}{2}, \,\,\,  &{\Lambda}_n[\ell,\ell] > \frac{\tilde{\mu}}{2} \\
0, \,\,\, &{\Lambda}_n[\ell,\ell] \leq \frac{\tilde{\mu}}{2}
\end{cases}.
\end{align}
Then, 
\begin{align}
\label{Eq_opt_soft_thr3}
\tilde{\Z}_n = \U_n \tilde{\bLambda}_n \V_n^H.
\end{align}

We turn to present an optimization scheme where \eqref{Eq_opt_prox_ell12} and \eqref{Eq_opt_prox_nuc} appear as sub-problems. 
Instead of the constrained form \eqref{Eq_opt_prob_convex}, we consider the related penalized form 
\begin{align}
\label{Eq_opt_prob_convex_pen}
&\minim{\{\Z_n\} \in \mathbb{C}^{N_\theta \times L}} \,\,\,  \beta \| [ \Z_1, \ldots, \Z_N ] \|_{1,2} + \mu \sum_{n=1}^N  \| \Z_n \|_* \nonumber\\
&\hspace{20mm} + \lambda \sum_{n=1}^{N} \sum_{\ell=1}^{L} \left \|\x_{\ell}(n) - \A_{\ell} \Z_n[:,\ell] \right \|_2^2,
\end{align}
where $\lambda$ is a positive hyper-parameter (in addition to $\beta,\mu$).
Define $\Z \triangleq [ \Z_1, \ldots, \Z_N ] \in \mathbb{C}^{N_\theta \times NL}$. 
Next, we apply variable splitting using an auxiliary variable $\G \in \mathbb{C}^{N_\theta \times NL}$ and a constraint $\G=\Z$.
Denote also 
\begin{align}
\label{Eq_opt_aux_def}
\G_n &\triangleq \G[:,(n-1)L+1:nL] \in \mathbb{C}^{N_\theta \times L}, \\ \nonumber
\g_{\ell,n} &\triangleq \G[:,(n-1)L+\ell] \in \mathbb{C}^{N_\theta \times 1}.
\end{align}
We have
\begin{align}
\label{Eq_opt_prob_convex_pen2}
&\minim{\Z,\G \in \mathbb{C}^{N_\theta \times NL}} \,\,\,  \beta \| \G \|_{1,2} + \mu \sum_{n=1}^N  \| \Z_n \|_* \nonumber\\
&\hspace{20mm} + \lambda \sum_{n=1}^{N} \sum_{\ell=1}^{L} \left \|\x_{\ell}(n) - \A_{\ell} \g_{\ell,n} \right \|_2^2 \nonumber\\
&\hspace{7mm} \mathrm{s.t.} \,\,\,\, \G=\Z
\end{align}
The new constrained problem \eqref{Eq_opt_prob_convex_pen2} can be optimized by ADMM (see the overview in \cite{boyd2011distributed}) as follows.
First, we construct the augmented Lagrangian with scaled dual variable $\Y \in \mathbb{C}^{N_\theta \times NL}$
\begin{align}
\label{Eq_opt_prob_aug_Lag}
&\mathcal{L}(\G,\Z,\Y)= \beta \| \G \|_{1,2} + \lambda \sum_{n=1}^{N} \sum_{\ell=1}^{L} \left \|\x_{\ell}(n) - \A_{\ell} \g_{\ell,n} \right \|_2^2 \nonumber\\
&+ \mu \sum_{n=1}^N  \| \Z_n \|_* + \rho \|\G-\Z+\Y\|_F^2 - \rho \|\Y\|_F^2.
\end{align}
Here, $\rho$ is the ADMM penalty hyper-parameter. 
Then, $\mathcal{L}(\G,\Z,\Y)$ is alternately optimized with respect to $\G,\Z$ and $\Y$. Specifically, initialize $\G^{(0)}=\Z^{(0)}=\Y^{(0)}=\mathbf{0}$, 
and repeat 
the following three steps for $k=1,2,\ldots$ until convergence
\begin{align}
\label{Eq_opt_admm1}
\G^{(k)} &= \argmin{\G \in \mathbb{C}^{N_\theta \times NL}} \,\,\, \beta \| \G \|_{1,2} + \lambda \sum_{n=1}^{N} \sum_{\ell=1}^{L} \left \|\x_{\ell}(n) - \A_{\ell} \g_{\ell,n} \right \|_2^2 \nonumber \\ & ~~~~~~~~~~~~~~~~~~ + \rho \|\G - \Z^{(k-1)} + \Y^{(k-1)} \|_F^2, \\
\label{Eq_opt_admm2}
\Z_n^{(k)} &= \argmin{\Z_n \in \mathbb{C}^{N_\theta \times L}} \,\,\, \mu \| \Z_n \|_*  +  \rho \| \Z_n - \G_n^{(k)} - \Y_n^{(k-1)} \|_F^2, \nonumber \\ 
&\hspace{25mm} n=1,\ldots, N,  \\
\label{Eq_opt_admm3}
\Y^{(k)} &= \Y^{(k-1)} + \G^{(k)} - \Z^{(k)}.
\end{align}
Note that \eqref{Eq_opt_admm3} is a simple update and that the solution of \eqref{Eq_opt_admm2} is given by \eqref{Eq_opt_soft_thr1}-\eqref{Eq_opt_soft_thr3} with $\tilde{\mu}=\mu/\rho$ and $\tilde{\G}_n = \G_n^{(k)} + \Y_n^{(k-1)}$.

To optimize the sub-problem \eqref{Eq_opt_admm1} we apply FISTA \cite{beck2009fast}, which integrates the classical proximal gradient method with Nesterov's accelerated gradient \cite{nesterov1983method}. Specifically, we initialize $\tilde{\G}^{(0)}=\overline{\G}^{(1)}=\G^{(k-1)}$ (i.e., with $\G^{(k-1)}$ from the previous ADMM iteration) and $t_1=1$ and repeat the following steps for $q=1,2,\ldots$ until convergence of the sequence $\{ \tilde{\G}^{(q)} \}$
\begin{align}
\label{Eq_opt_fista1}
\tilde{\G}^{(q)} &= \argmin{\G \in \mathbb{C}^{N_\theta \times NL}} \,\,\,  \beta \gamma \| \G \|_{1,2} + \| \G - ( \overline{\G}^{(q)} - \gamma \Delta \overline{\G}^{(q)} ) \|_F^2, \\
t_{q+1} &= \frac{ 1+\sqrt{1+4t_q^2} }{2}, \\
\label{Eq_opt_fista2}
\overline{\G}^{(q+1)} &= \tilde{\G}^{(q)} + \frac{t_q-1}{t_{q+1}} (\tilde{\G}^{(q)}-\tilde{\G}^{(q-1)}),
\end{align}
where $\gamma$ is a step-size and $\Delta \overline{\G}^{(q)}$ denotes the Wirtinger gradient\footnote{Informally, Wirtinger derivative of a real-valued function with respect to a complex variable treats the complex conjugate of the variable as a constant. For its usage in array processing, see \cite{brandwood1983complex,kreutz2009complex}.} of the last two (continuously differentiable) terms of \eqref{Eq_opt_admm1}, i.e.,  
$f(\G) = \lambda \sum_{n,\ell=1}^{N,L} \left \|\x_{\ell}(n) - \A_{\ell} \g_{\ell,n} \right \|_2^2 + \rho \|\G - \Z^{(k-1)} + \Y^{(k-1)} \|_F^2$,
at the point $\overline{\G}^{(q)}$, and is given by
\begin{align}
\label{Eq_opt_fista_grad}
\Delta \overline{\G} &= \lambda [ \Delta \overline{\g}_{1,1} , \ldots , \Delta \overline{\g}_{L,1}, \ldots,  \Delta \overline{\g}_{L,N} ] \nonumber \\
&\hspace{5mm} + \rho ( \overline{\G} - \Z^{(k-1)} + \Y^{(k-1)} ) \nonumber \\
\Delta \overline{\g}_{\ell,n} & \triangleq \A_{\ell}^H \left ( \A_{\ell} \overline{\G}[:,(n-1)L+\ell] - \x_{\ell}(n) \right ).
\end{align}
Note that \eqref{Eq_opt_fista1} has a closed-form solution, given by \eqref{Eq_opt_prox_ell12_solution} with $\tilde{\beta}=\beta \gamma$ and $\tilde{\Z}=\overline{\G}^{(q)} - \gamma \Delta \overline{\G}^{(q)}$.
As for the step-size, for simplification we use a constant step-size of $\gamma =  1/(\lambda \cdot \mathrm{max}_{\ell}\|\A_{\ell}^H\A_{\ell}\| + \rho)$, rather than line search or backtracking. Here, $\|\A_{\ell}^H\A_{\ell}\|$ is the spectral norm of $\A_{\ell}^H\A_{\ell}$, or equivalently, its largest eigenvalue, which can be computed extremely fast by the power method.
Note that $1/\gamma$ upper bounds the Lipschitz constant of the gradient of $f(\G)$ (the smooth part in \eqref{Eq_opt_admm1}), and thus (accelerated) convergence of FISTA to the global minimum is ensured \cite{beck2009fast}.

The high-level description of the optimization scheme, with additional implementation details, is presented in Algorithm~\ref{opt_alg}. 
Empirically, the stopping criteria on 
the residuals $r_{in}$ and $r_{out}$ happen much before the strict conditions on the maximum iterations. We verfied that Algorithm~\ref{opt_alg} yields the same optimal value (and also the same minimizer) as obtained when \eqref{Eq_opt_prob_convex_pen} is solved using the CVX Matlab package. Yet, Algorithm~\ref{opt_alg} is faster than CVX by several orders. The reason is that CVX calls a standard interior-point SDP solver that scales polynomially with $N$, $N_\theta$ and $L$.
On the other hand, Algorithm~\ref{opt_alg} requires only first-order derivatives of smooth terms, uses analytical solutions for non-smooth terms, and scales only linearly with the number of snapshots $N$. 
\tomt{For example, using Intel-i7 laptop, the typical runtime values of the solvers for our experiments with $N=5$ snapshots (detailed in the sequel) are: 
520s, 560s, and 3s, for CVX applied on \eqref{Eq_opt_prob_convex}, CVX applied on \eqref{Eq_opt_prob_convex_pen}, and Algorithm~\ref{opt_alg} (constructed for \eqref{Eq_opt_prob_convex_pen}), respectively.
}

\begin{algorithm}[t]
\caption{Integrated ADMM + FISTA scheme for minimizing \eqref{Eq_opt_prob_convex_pen}}
\vspace{2mm}
\kwInput{$ \{ \x_\ell(n) \}_{n,\ell=1}^{N,L}$, $\{ \A_{\ell} \}_{\ell=1}^{L}$, $\beta$, $\mu$, $\lambda$.}
\kwDefaults{$\rho=10$, $\gamma =  1/(\lambda \cdot \mathrm{max}_{\ell}\|\A_{\ell}^H\A_{\ell}\| + \rho)$, $\G^{(0)}=\Z^{(0)}=\Y^{(0)}=\mathbf{0}$.}
\Do{$k<$250 and $r_{out}>$5e-6}{
    $k = k+1$\;
    $q = 0$\;
    $\tilde{\G}^{(0)}=\overline{\G}^{(1)}=\G^{(k-1)}$\;
    \Do{$q<$1000 and $r_{in}>$5e-6}{
    $q = q+1$\;
    Compute $\tilde{\G}^{(q)}$ using \eqref{Eq_opt_fista1}-\eqref{Eq_opt_fista2} (utilizing \eqref{Eq_opt_prox_ell12_solution} with $\tilde{\beta}=\beta \gamma$ and $\tilde{\Z}=\overline{\G}^{(q)} - \gamma \Delta \overline{\G}^{(q)}$)\;
    $r_{in} \leftarrow \|\tilde{\G}^{(q)}-\tilde{\G}^{(q-1)}\|_F / \| \tilde{\G}^{(q-1)} \|_F$\;
    }
    $\G^{(k)} \leftarrow \tilde{\G}^{(q)}$\;
    \For{$n$ from 1 to $N$}{
        Compute $\Z_n^{(k)}$ using  \eqref{Eq_opt_soft_thr1}-\eqref{Eq_opt_soft_thr3} with $\tilde{\mu}=\mu/\rho$ and $\tilde{\G}_n = \G_n^{(k)} + \Y_n^{(k-1)}$\;
    }
    $\Z^{(k)} \leftarrow [ \Z_1^{(k)}, \ldots, \Z_N^{(k)} ]$\;
    $\Y^{(k)} \leftarrow \Y^{(k-1)} + \G^{(k)} - \Z^{(k)}$\;
    $r_{out} \leftarrow \| \G^{(k)} - \Z^{(k)} \|_F / \| \Z^{(k)} \|_F$\;
}    
\kwOutput{$\{\hat{\Z}_n \} = \{ \Z_n^{(k)}\}_{n=1}^N$.}
\label{opt_alg}
\end{algorithm}


\section{Numerical Results}
\label{Sec:EXPERIMENTS}

In this section, 
we perform 
computer simulations 
to evaluate the performance of the two proposed DOA estimation methods.
The strategy where we directly estimate the DOAs from $\{\hat{\Z}_n\}$ will be referred to as {\em Proposed1}.
The strategy where we estimate the phase shifts from $\{\hat{\Z}_n\}$ and then apply (quasi-)coherent 
DOA estimation methods 
will be referred to as {\em Proposed2}.

In experiments with a single snapshot ($N=1$) we solve \eqref{Eq_opt_prob_convex} using CVX, with hyper-parameters $\beta=0.1$, $\mu=0.9$ and $C=2$, and 
in {\em Proposed2} we minimize \eqref{Eq_opt_prob_convex_phase_corrected} with $C=2$ (after the phase-correction). 
In experiments with multiple snapshots (namely, $N=5, 25$), 
CVX has been shown to be very slow. Thus, 
we solve \eqref{Eq_opt_prob_convex_pen} via the much faster  Algorithm~\ref{opt_alg}, with hyper-parameters $\beta=0.1$, $\mu=0.9$ and $\lambda=1/\sqrt{2\sigma^2\mathrm{ln}(5M)} \cdot 1/M$ (recall that $\sigma^2$ is the noise variance).\footnote{We note that Algorithm~\ref{opt_alg} can be applied also for $N=1$, providing results that are very similar to those obtained when solving \eqref{Eq_opt_prob_convex} using CVX (with the stated hyper-parameter settings).}  
The first factor in $\lambda$ is similar to the settings used in classical papers on $\ell_1$-norm sparsity methods \cite{chen1994basis,chen2001atomic,donoho1995noising}, and we multiply it by $1/M$ because in these papers the columns of the measurement matrix are scaled to unit Euclidean norm. 
The second stage of {\em Proposed2} for multiple snapshots includes the classical MUSIC algorithm \cite{schmidt1986multiple} with forward-backward smoothing \cite{pillai1989forward} (which is faster and simpler than $\ell_1$-norm sparsity methods but cannot be applied for $N=1$).

We compare our strategies with several reference methods: 

(a) ``Non-coherent" MUSIC, which
treats the $N$ snapshots of each of the $L$ sub-arrays as $NL$ snapshots of a single sub-array (and thus does not require estimating $\{ \phi_\ell(n) \}$). We use this method with forward-backward smoothing \cite{pillai1989forward} (within each sub-array) that empirically improved its performance. Applying this non-coherent MUSIC is possible since in our experiments the sub-arrays are uniform linear arrays  (ULAs) 
with similar sensor configuration. 

(b) The alternating minimization method, termed {\em Joint-AM} in \cite{tirer2020method}, which is based on non-convex optimization of the likelihood function.
Note that, unlike convex optimization methods, {\em Joint-AM} is strongly affected by its initialization. 
To enhance the method's performance in \cite{tirer2020method}, at the price of heavy computational complexity, {\em Joint-AM} is repeatedly applied there as follows. First, it is applied with initialization that is obtained by a (non-coherent processing) variant of \cite{ziskind1988maximum} and yields estimated DOAs. Then, a ``multiple initializations trick" is carried out: many more initializations are produced from perturbations of the previously estimated DOAs and {\em Joint-AM} is repeatedly applied. 
For fair comparison, here we apply {\em Joint-AM} without the multiple initializations trick used in \cite{tirer2020method}, which makes it very slow. Even now, it is about 3-5 times slower than {\em Proposed2}. 
Note also that one can use the results of our approach to (better) initialize the non-convex optimization considered in \cite{tirer2020method}.

(c) Methods that are similar to our {\em Proposed1}, but use only the sparsity promoting term (by setting $\beta=1$ and $\mu=0$) or only the low-rankness promoting term (by setting $\beta=0$ and $\mu=1$) in the convex program.
The other configurations remain the same, including the final rank-1 approximation step of Section~\ref{sec:Prop1}.
These methods will be referred to as {\em SparsityOnly} and {\em LowRankOnly}. 
Note that 
{\em SparsityOnly} 
resembles an adaptation of the self-calibration algorithms \cite{ling2015self,hung2017low} to the considered non-coherent problem. (The final rank-1 approximation appears only in \cite{hung2017low} but it is shown there to yield better results than \cite{ling2015self}).

Apart from the above reference methods, we also display the Cramér-Rao lower bound (CRLB) for the considered  problem \eqref{Eq_subarray_model}, which is derived in \cite{tirer2020method}.

Note that all the examined methods require a DOA grid (containing different hypotheses of a scalar DOA). Specifically, in our experiments we 
use DOA grid between $-45^\circ$ and $45^\circ$ with resolution of $\Delta = 0.1^\circ$
for all the methods. 
Note 
also
that all the examined methods, except {\em Joint-AM}, are based on estimating the DOAs as the peaks of $N_\theta \times 1$ ``spectrums"  
over the DOA grid. We determine the estimated DOA vector $\hat{\btheta}$ of each method according to its spectrum's $Q$ highest peaks (local maxima), where $Q$ is the true number of sources. 
If the number of peaks in the spectrum 
is smaller than $Q$ then the remaining DOA estimates are taken according to the maximal spectrum values (even if these points are not local maxima). 

To obtain statistical results, we perform $N_{\text{exp}}=250$ Monte Carlo experiments at each signal-to-noise ratio (SNR) between 0 dB to 30 dB in every examined scenario (where the SNR is defined per antenna). 
In each Monte Carlo experiment, for each of the snapshots $n=1,\ldots,N$ we draw complex Gaussian noise $\{\e_{\ell}(n)\}$, complex Gaussian signal $\tilde{\s}(n)$ with equal power for all sources, and phase shifts $\{\phi_{\ell}(n)\}$ that are uniformly distributed over $[0, 2\pi]$. 

\tomtr{
In practice, 
the unknown true DOAs come from a continuous interval of angles, and thus (with probability one) do not exactly fall on grid points of the estimators. 
We take this into account using 
a {\em perturbation methodology}: 
another source of randomness in the Monte Carlo experiments that comes from slightly perturbing the ground truth DOAs. 
Specifically, in
each scenario, instead of fixing a DOA to an angle $\tilde{\theta}_{gt}$, in each Monte Carlo experiment 
we draw the true DOA $\theta_{gt}$ from the uniform distribution $U[\tilde{\theta}_{gt}-\frac{\Delta}{2}, \tilde{\theta}_{gt}+\frac{\Delta}{2}]$ 
(recall that $\Delta$ is the estimators' grid resolution).
Thus, it is ensured that no algorithm can have any advantage that originates from the grid quantization. 
To simplify the presentation in this section, when we state that the sources at an examined scenario are at $\sim \tilde{\btheta}_{gt}$ this means that in each Monte Carlo experiment they are drawn from $U[\tilde{\btheta}_{gt}-\frac{\Delta}{2}, \tilde{\btheta}_{gt}+\frac{\Delta}{2}]$.}

The performance of each method is measured by the root mean square error (RMSE), 
which is defined by
\begin{align}
\label{Eq_rmse}
\text{RMSE} = \sqrt{ \frac{1}{N_{\text{exp}} \cdot Q}\sum \limits_{i=1}^{N_{\text{exp}}} \| \hat{\btheta}(i) - \btheta_{gt}(i) \|^2 },
\end{align}
where $\hat{\btheta}(i)$ is the estimated DOA vector at the $i$-th experiment and \tomtr{$\btheta_{gt}(i)$ is the true DOA vector of the $i$-th experiment, which, as explain above, is a (slightly) randomly perturbed version of the fixed $\tilde{\btheta}_{gt}$ associated with the examined scenario. 
Note that in our experiments the RMSE is lower bounded by
the standard deviation of the
quantization error $\sqrt{\frac{\Delta^2}{12}} = \sqrt{\frac{0.1^2}{12}} \approx 0.029^\circ$, which is the standard deviation of a uniformly distributed random variable with support length of $\Delta = 0.1^\circ$. 
Yet, we have chosen the grid resolution $\Delta$ to be small enough 
such that the grid quantization error is sufficiently smaller than the estimators' RMSE at 
the examined scenarios and SNRs. 
Hence the degradation due to quantization error is not significant and 
does not affect the comparison of the methods.\footnote{In general, if one wishes to use a very small $\Delta$ for improved resolution, then multi-resolution grid search strategies can be applied to reduce the runtime (e.g., repeatedly apply the algorithm with grid of higher resolution in the vicinity of the previously estimated DOAs).}}

\begin{figure}[t]
  \centering
  \begin{subfigure}[b]{0.9\linewidth}
    \centering\includegraphics[width=220pt]{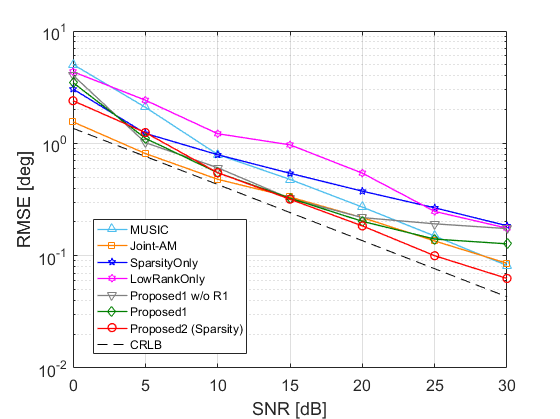} 
  \end{subfigure}%
  \caption{DOA RMSE vs. SNR for sources at 
  $\sim \{ 0^\circ, 15^\circ \}$,
  using 4 uniform linear sub-arrays, each of 6 elements, and $N=1$ snapshot.}
\label{fig:Q2_L4_M24_new2}

\vspace{5mm}

  \centering
  \begin{subfigure}[b]{0.9\linewidth}
    \centering\includegraphics[width=220pt]{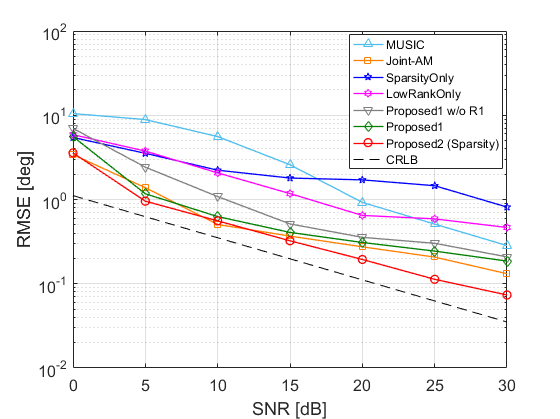}
  \end{subfigure}%
  \caption{DOA RMSE vs. SNR for sources at 
  $\sim \{-15^\circ, 0^\circ, 15^\circ, 30^\circ \}$,
  using 4 uniform linear sub-arrays, each of 6 elements, and $N=1$ snapshot.}
\label{fig:Q4_L4_M24_new}
\end{figure}

\begin{figure}
  \centering
  \begin{subfigure}[b]{0.49\linewidth}
    \centering\includegraphics[width=120pt]{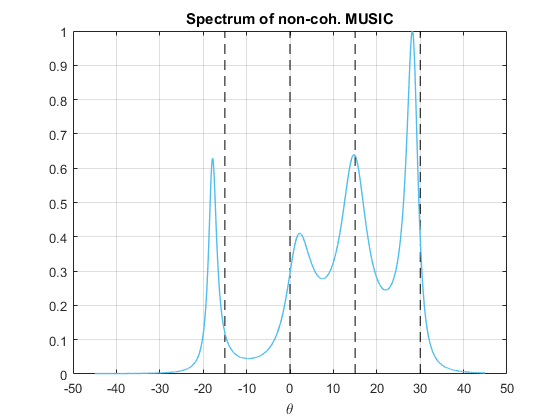}     
  \end{subfigure}%
  \begin{subfigure}[b]{0.49\linewidth}
    \centering\includegraphics[width=120pt]{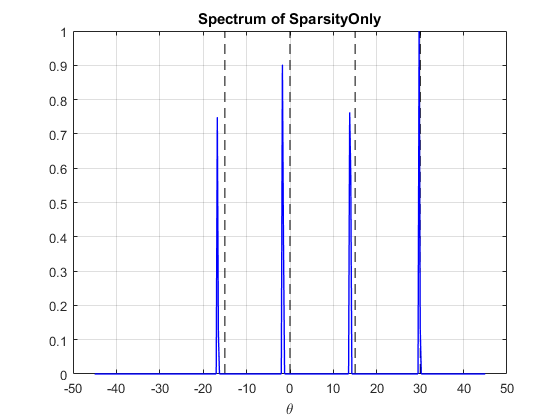}    
  \end{subfigure}%
\\
  \begin{subfigure}[b]{0.49\linewidth}
    \centering\includegraphics[width=120pt]{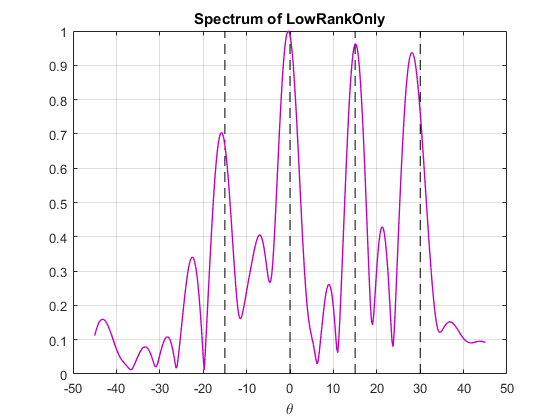}
  \end{subfigure}
  \begin{subfigure}[b]{0.49\linewidth}
    \centering\includegraphics[width=120pt]{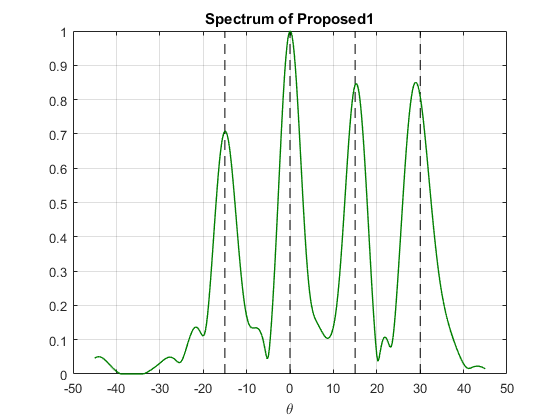}    
  \end{subfigure}
\\
  \begin{subfigure}[b]{0.49\linewidth}
    \centering\includegraphics[width=120pt]{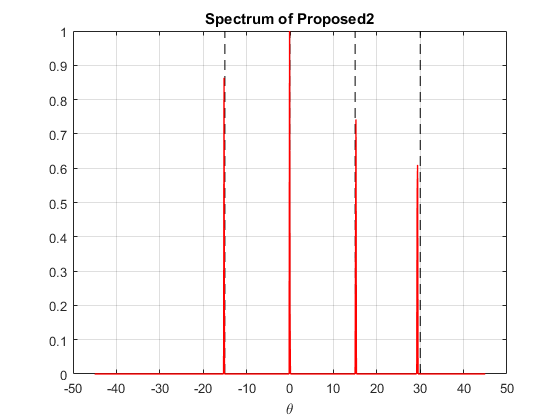}    
  \end{subfigure}
  \caption{The spectrums of the different methods for one realization at SNR of 10 dB, for sources at $-15^\circ$, $0^\circ$, $15^\circ$ and $30^\circ$, using 4 uniform linear sub-arrays, each of 6 elements, and $N=1$ snapshot. From left to right and top to bottom: {\em non-coherent-MUSIC}, {\em SparsityOnly}, {\em LowRankOnly}, {\em Proposed1}, and {\em Proposed2}. The true DOAs are marked by vertical dashed black lines. 
  }
\label{fig:spectrums}
\end{figure}

\subsection{Single Snapshot Experiments}

We start with single snapshot scenarios, i.e., $N=1$. 
The requirement to estimate the DOAs from a single snapshot is encountered 
in many practical applications, e.g.,~when the sources and/or receivers are moving. 
We consider 
a ULA of $M=24$ elements, with half wavelength spacing, partitioned into $L=4$ sub-arrays of size $M_{\ell}=6$ each. 
We first consider the case of
$Q=2$ sources 
located at angles 
$\sim \{ 0^\circ, 15^\circ \}$
(the boresight angle of the entire array is $0^\circ$).  
The DOA RMSE results are presented in Fig.~\ref{fig:Q2_L4_M24_new2}. 

From Fig.~\ref{fig:Q2_L4_M24_new2} it
can be seen that {\em Proposed1}, which promotes {\em both} sparsity and  low-rankness {\em within} the convex optimization, is better than {\em SparsityOnly} and {\em LowRankOnly}. 
In fact, {\em SparsityOnly} and {\em LowRankOnly} are even inferior to the non-coherent MUSIC. 
\tomtr{
Regarding {\em Proposed1}, we also present the results of its variant, denoted by ``Proposed1 w/o R1", which does not use the rank-1 approximation step that is described in Section~\ref{sec:Prop1}. Comparing the two shows the advantage of using this step at high SNR.} 
As for {\em Proposed2}, in the considered experiment it 
is competitive with {\em Joint-AM}. 
The better performance of {\em Proposed2} over {\em Proposed1} (especially at high SNR) demonstrates the advantage of using the solution of the joint sparse and low-rank optimization problem for estimating the phase shifts rather than the DOAs.

Spectrum-based methods (such as our proposed strategies) are especially preferable 
over maximum likelihood techniques (such as {\em Joint-AM}) 
when the number of sources is large. 
Thus, we turn to examine a scenario with $Q=4$ sources, located at angles $\sim \{ -15^\circ, 0^\circ, 15^\circ, 30^\circ \}$. The rest of the configuration remains as before (e.g., we still consider $N=1$ snapshot). 
The DOA RMSE results are presented in Fig.~\ref{fig:Q4_L4_M24_new}, and the estimators' spectrums for one realization at SNR of 10 dB are displayed in Fig.~\ref{fig:spectrums}. 

Comparing the spectrums of {\em Proposed1} and {\em SparsityOnly} in Fig.~\ref{fig:spectrums}, we see that the nuclear norm that is used in {\em Proposed1} reduces the sparsity but significantly improves the accuracy of the DOAs estimates (i.e., the location of the peaks).
Similarly, comparing the spectrums of {\em Proposed1} and {\em LowRankOnly} shows that the $\ell_{1,2}$-norm that is used in {\em Proposed1} both reduces 
false peaks 
and improves the accuracy. 
From Fig.~\ref{fig:spectrums}
it also observed that the non-coherent MUSIC is more prone to missing sources that the other methods at low to medium SNR, which is inline with the results in Fig.~\ref{fig:Q4_L4_M24_new}. 

\tomtr{
Note that Fig.~\ref{fig:Q4_L4_M24_new} presents the results of {\em Proposed1} both with and without the rank-1 approximation step, and demonstrates that this step indeed improves the results and leads to the fact that {\em Proposed1} is competitive with {\em Joint-AM}.} 
From Fig.~\ref{fig:Q4_L4_M24_new} 
it is also observed that 
{\em Proposed2} outperforms all the examined methods. 
\tomt{
Comparing Figs.~\ref{fig:Q2_L4_M24_new2} and \ref{fig:Q4_L4_M24_new} 
shows that the gap between the proposed methods (and the CRLB) and the non-coherent MUSIC is increased for $Q=4$ sources compared to $Q=2$ sources. 
This is aligned with the results and analysis in \cite{tirer2020method}, which imply that compromising on non-coherent processing (rather than using more complex methods to utilize
the phase information between different sub-arrays) 
is less worthy as the number of sources grows.}

\begin{figure}[t]
  \centering
  \begin{subfigure}[b]{0.9\linewidth}
    \centering\includegraphics[width=220pt]{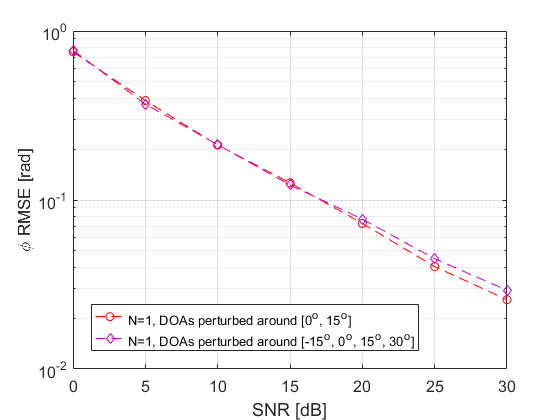}
  \end{subfigure}%
  \caption{Phase shifts RMSE vs. SNR for sources at $\sim \{ 0^\circ, 15^\circ \}$ and sources at $\sim \{ -15^\circ, 0^\circ, 15^\circ, 30^\circ \}$, using 4 uniform linear sub-arrays, each of 6 elements, and $N=1$ snapshot.}
\label{fig:phaseShifts_N1_Q2_spread_Q4_spread_M24_L4}
\end{figure}

\begin{figure}
  \centering
  \begin{subfigure}[b]{0.9\linewidth}
    \centering\includegraphics[width=220pt]{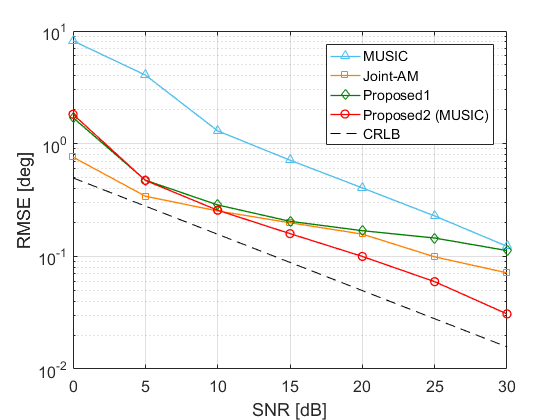} 
  \end{subfigure}%
  \caption{DOA RMSE vs. SNR for sources at 
  $\sim \{ -15^\circ, 0^\circ, 15^\circ, 30^\circ \}$, 
  using 4 uniform linear sub-arrays, each of 6 elements, and $N=5$ snapshots.}
\label{fig:Q4_L4_M24_N5}

\vspace{5mm}

  \centering
  \begin{subfigure}[b]{0.9\linewidth}
    \centering\includegraphics[width=220pt]{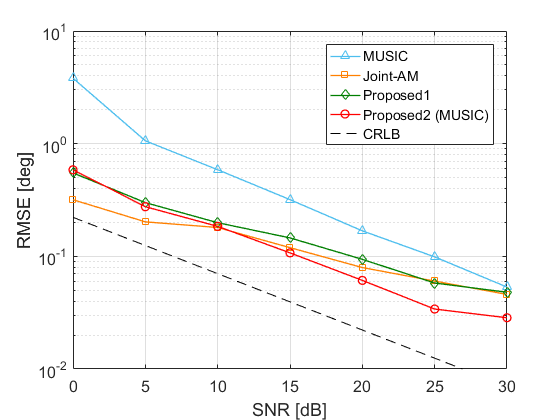} 
  \end{subfigure}%
  \caption{DOA RMSE vs. SNR for sources at 
  $\sim \{ -15^\circ, 0^\circ, 15^\circ, 30^\circ \}$, 
  using 4 uniform linear sub-arrays, each of 6 elements, and $N=25$ snapshots.} 
\label{fig:Q4_L4_M24_N25}
\end{figure}

Next, we present in Fig.~\ref{fig:phaseShifts_N1_Q2_spread_Q4_spread_M24_L4} the RMSE of the sub-arrays' phase shifts estimates, which are used in {\em Proposed2}, for the two 
different DOA scenarios in Figs.~\ref{fig:Q2_L4_M24_new2} and \ref{fig:Q4_L4_M24_new}. 
We note that our RMSE calculation for the phase offsets estimates was invariant to $2\pi$ shifts as well as to a constant shift in each $L$-tuple $\{ \hat{\phi}_\ell(n) \}_{\ell=1}^{L}$ (i.e., a constant shift in the phase offsets of all the sub-arrays in a given snapshot). These invariances do not affect the DOA estimation after the ``phase correction" step. 
It can be seen that despite the potential difficulty of having more sources, the 
quality of the sub-arrays' phase shifts estimation 
is similar in both cases.

\begin{figure}
  \centering
  \begin{subfigure}[b]{0.9\linewidth}
    \centering\includegraphics[width=220pt]{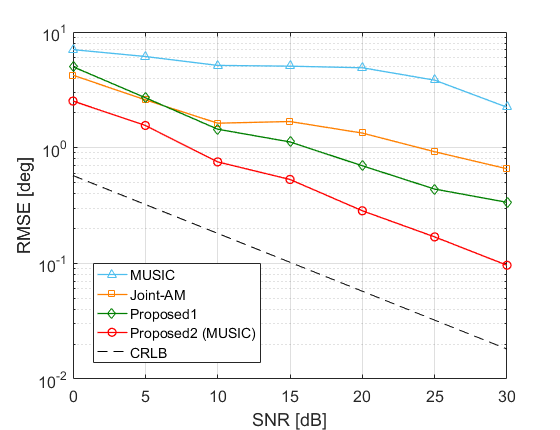}
  \end{subfigure}%
  \caption{DOA RMSE vs. SNR for sources at 
  $\sim \{ -7.5^\circ, 0^\circ, 7.5^\circ, 15^\circ \}$, 
  using 4 uniform linear sub-arrays, each of 6 elements, and $N=5$ snapshots.}
\label{fig:Q4_L4_M24_N5_dense}

\vspace{5mm}

  \centering
  \begin{subfigure}[b]{0.9\linewidth}
    \centering\includegraphics[width=220pt]{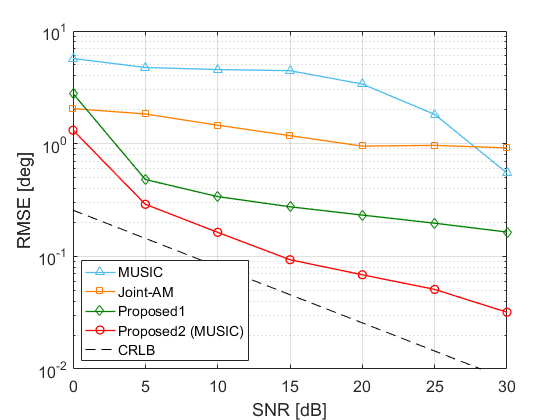}
  \end{subfigure}%
  \caption{DOA RMSE vs. SNR for sources at 
  $\sim \{ -7.5^\circ, 0^\circ, 7.5^\circ, 15^\circ \}$, 
  using 4 uniform linear sub-arrays, each of 6 elements, and $N=25$ snapshots. }
\label{fig:Q4_L4_M24_N25_dense}
\end{figure}

\subsection{Multiple Snapshots Experiments}


We turn now to examine multiple snapshots scenarios, specifically, $N=\{5,25\}$ snapshots, still with $Q=4$ sources located at $\sim \{ -15^\circ, 0^\circ, 15^\circ, 30^\circ \}$. 
To simplify the presentation, we do not display the results of {\em SparsityOnly} and {\em LowRankOnly} that are consistently inferior to those of {\em Proposed1}. 
Recall that in the multiple snapshots experiments we use Algorithm~\ref{opt_alg} to obtain the matrices $\{ \hat{\Z}_n \}$, as we observed that general-purpose solvers are extremely slow. 
The DOA RMSE results are presented in Figs.~\ref{fig:Q4_L4_M24_N5} and \ref{fig:Q4_L4_M24_N25}. 

It can be seen that starting from a quite low SNR {\em Proposed1} is competitive with {\em Joint-AM}, and that {\em Proposed2} outperforms the other methods. All the methods exhibit performance gain compared to the single snapshot case.

Next, we examine again scenarios with $N=\{5,25\}$ snapshots, but this time with denser $Q=4$ sources, located at $\sim \{ -7.5^\circ, 0^\circ, 7.5^\circ, 15^\circ \}$, which further complicated the estimation problem. The rest of the configuration remains as before. 
The DOA RMSE results are presented in Figs.~\ref{fig:Q4_L4_M24_N5_dense} and \ref{fig:Q4_L4_M24_N25_dense}.

In this challenging setting the performance of the non-coherent MUSIC significantly deteriorates as it fails to resolve the dense sources until very high SNR values.
The performance of {\em Joint-AM} is also deteriorated as its non-convex optimization is more likely of getting trapped in bad local minima.
Indeed, our {\em Proposed1} outperforms {\em Joint-AM} in these experiments.
Once again, the best results are obtained by our {\em Proposed2}, which uses the convex optimization to estimate the sub-arrays' phase shifts rather than then directly the DOAs.

\begin{figure}[t]
  \centering
  \begin{subfigure}[b]{0.9\linewidth}
    \centering\includegraphics[width=220pt]{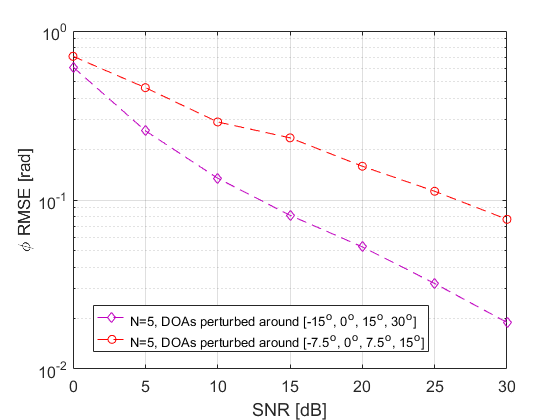}
  \end{subfigure}%
  \caption{Phase shifts RMSE vs. SNR for sources at $\sim \{ -15^\circ, 0^\circ, 15^\circ, 30^\circ \}$ and sources at $\sim \{ -7.5^\circ, 0^\circ, 7.5^\circ, 15^\circ \}$, using 4 uniform linear sub-arrays, each of 6 elements, and $N=5$ snapshots.}
\label{fig:phaseShifts_N5_Q4_spread_Q2_dense_M24_L4}
\end{figure}

Finally, we present in Fig.~\ref{fig:phaseShifts_N5_Q4_spread_Q2_dense_M24_L4} the RMSE of the estimated phase shifts obtained in {\em Proposed2} for $N=5$ snapshots for sources at $\sim \{ -15^\circ, 0^\circ, 15^\circ, 30^\circ \}$ and at $\sim \{ -7.5^\circ, 0^\circ, 7.5^\circ, 15^\circ \}$ (associated with Figs.~\ref{fig:Q4_L4_M24_N5} and \ref{fig:Q4_L4_M24_N5_dense}).
As expected, in the tougher scenario (associated with Fig.~\ref{fig:Q4_L4_M24_N5_dense}) the performance of the proposed phase estimation is decreased. Yet, it still exhibits an appealing monotonic improvement as the SNR increases.

\subsection{The Methods Runtime}

\tomtr{
We finish the experiments section with a comparison of the average runtime values of the different methods in different scenarios (which differ in the number of snapshots) with Intel-i7 laptop. 
Table~\ref{table:Runtime} shows the runtimes of the non-coherent MUSIC, {\em Joint-AM}, and the two proposed methods.
Specifically, for {\em Proposed1} we present the runtime both with Algorithm~\ref{opt_alg} (the ADMM + FISTA scheme) and with CVX for solving \eqref{Eq_opt_prob_convex_pen} and \eqref{Eq_opt_prob_convex}, respectively. 
The runtime results of {\em Proposed2} are obtained with Algorithm~\ref{opt_alg} for solving \eqref{Eq_opt_prob_convex_pen} and either Algorithm~\ref{opt_alg_sync} or CVX for solving \eqref{Eq_opt_prob_convex_alpha}.} 

\tomtr{
From Table~\ref{table:Runtime} we see that 
{\em Proposed1} is significantly accelerated when using Algorithm~\ref{opt_alg} instead of the CVX implementation. The table also shows that when both {\em Proposed1} and {\em Proposed2} use Algorithm~\ref{opt_alg}, and \eqref{Eq_opt_prob_convex_alpha} in {\em Proposed2} is solved per snapshot via CVX, then the additional runtime of {\em Proposed2} (with MUSIC in the second stage) is relatively small (about 33\%). The CVX implementation of \eqref{Eq_opt_prob_convex_alpha} has relatively small runtime because the dimension of the optimization variable $\V \in \mathbb{C}^{L \times L}$ in \eqref{Eq_opt_prob_convex_alpha} is small (recall that $L$ is the number of sub-arrays). Nevertheless, the increase in {\em Proposed2} runtime is almost completely reduced when \eqref{Eq_opt_prob_convex_alpha} is solved with Algorithm~\ref{opt_alg_sync}. Finally, comparing the runtimes of all the different methods we observe that, unsurprisingly, the non-coherent MUSIC is by far the fastest method. Yet, recall that its performance is significantly inferior to the others in terms of accuracy. One can also see that both {\em Proposed1} and {\em Proposed2} with Algorithms~\ref{opt_alg} and \ref{opt_alg_sync} are faster than the {\em Joint-AM} reference method.}

\begin{table*}
\centering
    \caption{The Average Runtime of the Methods in Different Scenarios with Intel-i7 Laptop} 
    \label{table:Runtime}
    \begin{tabular}{ | c | c | c | c | c | c | c |}
    \hline
        & MUSIC
        & Joint-AM
        & \begin{tabular}{@{}c@{}}Proposed1 \\ with Alg 1 \end{tabular}
        & \begin{tabular}{@{}c@{}}Proposed1 \\  with CVX \end{tabular}
        & \begin{tabular}{@{}c@{}}Proposed2 \\ with Alg 1 and Alg 2 for \eqref{Eq_opt_prob_convex_alpha} \end{tabular}
        & \begin{tabular}{@{}c@{}} Proposed2 \\ with Alg 1 and CVX for \eqref{Eq_opt_prob_convex_alpha} \end{tabular} \\ \hline
\begin{tabular}{@{}c@{}}Scenario of Fig.~\ref{fig:Q4_L4_M24_new} \\ $N=1$ snapshot \end{tabular}
        & $<$ 0.02s & $\sim$ 8s & $\sim$ 0.6s & $\sim$ 30s &
        \begin{tabular}{@{}cc@{}}
        $\sim$ 2.6s &
        \begin{tabular}{@{}l@{}l@{}}
        \textcolor{gray}{$\sim$ 0.6s (Alg 1)} \\ \textcolor{gray}{+ 0.02s (1$\times$Alg2)} \\ \textcolor{gray}{+ 2s (CVX sparsity)}
        \end{tabular} 
        \end{tabular}
        & 
        \begin{tabular}{@{}cc@{}}
        $\sim$ 2.8s &
        \begin{tabular}{@{}l@{}l@{}}
        \textcolor{gray}{$\sim$ 0.6s (Alg 1)} \\ \textcolor{gray}{+ 0.2s (1$\times$CVX)} \\ \textcolor{gray}{+ 2s (CVX sparsity)}
        \end{tabular} 
        \end{tabular}
        \\ \hline
\begin{tabular}{@{}c@{}}Scenario of Fig.~\ref{fig:Q4_L4_M24_N5} \\ $N=5$ snapshots \end{tabular}
        & $<$ 0.02s & $\sim$ 10.5s & $\sim$ 3s & $\sim$ 520s &
        \begin{tabular}{@{}cc@{}}
        $\sim$ 3.1s &
        \begin{tabular}{@{}l@{}l@{}}
        \textcolor{gray}{$\sim$ 3s (Alg 1)} \\ \textcolor{gray}{+ 0.1s (5$\times$Alg2)} \\ \textcolor{gray}{+ 0.02s (MUSIC)}
        \end{tabular} 
        \end{tabular}
        & 
        \begin{tabular}{@{}cc@{}}
        $\sim$ 4s &
        \begin{tabular}{@{}l@{}l@{}}
        \textcolor{gray}{$\sim$ 3s (Alg 1)} \\ \textcolor{gray}{+ 1s (5$\times$CVX)} \\ \textcolor{gray}{+ 0.02s (MUSIC)}
        \end{tabular} 
        \end{tabular} 
        \\ \hline
\begin{tabular}{@{}c@{}}Scenario of Fig.~\ref{fig:Q4_L4_M24_N25} \\ $N=25$ snapshots \end{tabular}
        & $<$ 0.02s & $\sim$ 30s & $\sim$ 15s & \begin{tabular}{@{}c@{}} Computationally \\ impractical \end{tabular} & 
        \begin{tabular}{@{}cc@{}}
        $\sim$ 15.5s &
        \begin{tabular}{@{}l@{}l@{}}
        \textcolor{gray}{$\sim$ 15s  (Alg 1)} \\ \textcolor{gray}{+ 0.5s (25$\times$Alg2)} \\ \textcolor{gray}{+ 0.02s (MUSIC)}
        \end{tabular} 
        \end{tabular}
        & 
        \begin{tabular}{@{}cc@{}}
        $\sim$ 20s &
        \begin{tabular}{@{}l@{}l@{}}
        \textcolor{gray}{$\sim$ 15s (Alg 1)} \\ \textcolor{gray}{+ 5s (25$\times$CVX)} \\ \textcolor{gray}{+ 0.02s (MUSIC)}
        \end{tabular} 
        \end{tabular}
        \\ \hline        
    \end{tabular}
\end{table*}



\section{Conclusions}\label{sec:CONCLUSION}

We addressed the problem of estimating the DOAs of multiple sources for non-coherent sub-arrays that observe different unknown phase shifts at every snapshot. 
We formulated this problem as the reconstruction of joint sparse and low-rank matrices and solved the problem's convex relaxation.  
We designed an optimization scheme, based on integrating ADMM and FISTA, with complexity that scales with the number of snapshots much better than general-purpose solvers. 
Unlike methods that are based on maximum likelihood \cite{tirer2020method}, our approach can easily handle scenarios with a large number of sources and does not depend on the quality of its initialization. 

We showed two ways to estimate the DOAs. The first strategy estimates the DOAs directly from the solution of the convex problem and
outperforms non-coherent processing of the sub-arrays as well as other sparsity and low-rank based techniques.
The second strategy, which has shown even improved results and outperforms the method of \cite{tirer2020method},
uses the solution of the aforementioned convex problem for estimating the phase shifts rather than the DOAs, using another, computationally-light, convex relaxation that is practically tight. 
In our second strategy, the estimated phase shifts are used to ``correct" the observations, such that the DOA estimation can be done with plain methods of coherent arrays. 

\tomtb{The proposed approach achieves DOAs estimation with relatively high accuracy despite the non-coherency of the sub-arrays. 
As such, it can enable 
DOA estimation in systems with lower complexity and cost than 
\tomtb{those} 
that use fully coherent arrays.}


\appendix

\section{ADMM for synchronization}

\begin{algorithm}
\caption{ADMM scheme for optimizing \eqref{Eq_opt_prob_convex_alpha}}
\vspace{2mm}
\kwInput{$\Z_n$.}
\kwDefaults{$\rho=10$, $\gamma =  1/\rho$, 
$\Y^{(0)}=\mathbf{0}$, $\V^{(0)}=\tilde{\V}^{(0)} = \g\g^T$ where $\g \sim \mathcal{N}(\mathbf{0},\I_L)$.
}
\Do{$k<$250 and $r_{out}>$5e-6}{
    $k = k+1$\;
    $q = 0$\;
    $\V_0=\V^{(k-1)}$\;
    \Do{$q<$1000 and $r_{in}>$5e-6}{
    $q = q+1$\;
    Compute $\V_q$ using \eqref{App_Eq_opt_admm1_pgd}\;
    $r_{in} \leftarrow \|\V_q-\V_{q-1}\|_F / \| \V_{q-1} \|_F$\;
    }
    $\V^{(k)} \leftarrow \V_q$\;
    Compute $\tilde{\V}^{(k)}$ using \eqref{App_Eq_opt_admm2_solution}\;
    $\Y^{(k)} \leftarrow \Y^{(k-1)} + \V^{(k)} - \tilde{\V}^{(k)}$\;
    $r_{out} \leftarrow \| \V^{(k)} - \tilde{\V}^{(k)} \|_F / \| \tilde{\V}^{(k)} \|_F$\;
}    
\kwOutput{$\hat{\V}_n = \tilde{\V}^{(k)}$.}
\label{opt_alg_sync}
\end{algorithm}

\tomtr{
The convex problem in \eqref{Eq_opt_prob_convex_alpha} can be solved quite fast via general-purpose solvers when the number of sub-arrays $L$ is small. Yet, in this appendix we develop for it a dedicated first-order optimization scheme, which in our experiments is ten times faster. Restating the problem as a minimization task
\begin{align}
\label{App_Eq_opt_prob_convex_alpha}
&\minim{ \V \in \mathbb{C}^{L \times L}} \,\,\,  -\mathrm{Tr}( \hat{\Z}_n^H \hat{\Z}_n \V ) \nonumber\\
&\hspace{5mm} \mathrm{s.t.} \,\,\,\, \mathrm{diag}(\V)=\mathbf{1}_L, \,\, \V \succcurlyeq 0,
\end{align}
we see that its feasible set is the intersection of two convex sets, $\{\V : \mathrm{diag}(\V)=\mathbf{1}_L \}$ and $\{ \V : \V \succcurlyeq 0 \}$. Note that projecting a given $\V \in \mathbb{C}^{L \times L}$ onto each of the sets has a closed-form solution. Namely, projection onto the former set is done simply by setting the entries on the diagonal of $\V$ to 1, while projection onto the latter set is done by symmetrizing $\V$ (if it is non-symmetric) and setting its negative eigenvalues to zero (since positive semidefinite matrices have only non-negative eigenvalues). However, as far as we know, there is no closed-form solution for the projection onto the intersection of these sets, which rules out the use of a plain projected gradient descent scheme.}

\tomtr{
Following the above arguments, we choose to decouple the two convex sets by introducing an auxiliary variable $\tilde{\V} \in \mathbb{C}^{L \times L}$ and a constraint $\tilde{\V}=\V$, while expressing the previous two constraints with convex indicator functions. Formally, we have
\begin{align}
\label{App_Eq_opt_prob_convex_alpha2}
&\minim{ \V \in \mathbb{C}^{L \times L}} \,\,\,  -\mathrm{Tr}( \hat{\Z}_n^H \hat{\Z}_n \V ) \nonumber\\
&\hspace{15mm} + \mathbb{I}_{\{\V : \mathrm{diag}(\V)=\mathbf{1}_L \}}(\V) + \mathbb{I}_{\{ \tilde{\V} : \tilde{\V} \succcurlyeq 0 \}}(\tilde{\V}) \nonumber\\
&\hspace{5mm} \mathrm{s.t.} \,\,\,\, \tilde{\V}=\V,
\end{align}
where $\mathbb{I}_{\mathcal{C}}(\V)= \begin{cases} 
      0, & \V \in \mathcal{C} \\
      +\infty,  & \V \notin \mathcal{C}
   \end{cases}$,
which is 
the convex indicator function associated with the convex set $\mathcal{C}$. The new constrained problem can be optimized by ADMM as follows. First, we construct the augmented Lagrangian with the scaled dual variable $\Y \in \mathbb{C}^{L \times L}$
\begin{align}
\label{App_Eq_opt_prob_aug_Lag}
&\mathcal{L}(\V,\tilde{\V},\Y) = -\mathrm{Tr}( \hat{\Z}_n^H \hat{\Z}_n \V ) + \mathbb{I}_{\{\V : \mathrm{diag}(\V)=\mathbf{1}_L \}}(\V) \nonumber\\
&+ \mathbb{I}_{\{ \tilde{\V} : \tilde{\V} \succcurlyeq 0 \}}(\tilde{\V}) 
+ \rho \|\V - \tilde{\V} + \Y \|_F^2 - \rho \|\Y\|_F^2,
\end{align}
where $\rho$ is the ADMM penalty hyper-parameter. Then $\mathcal{L}(\V,\tilde{\V},\Y)$ is alternately optimized with respect to $\V,\tilde{\V}$ and $\Y$. Specifically, initialize 
$\Y^{(0)}=\mathbf{0}$ and $\V^{(0)}=\tilde{\V}^{(0)} = \g\g^T$ where $\g \sim \mathcal{N}(\mathbf{0},\I_L)$, 
and repeat the following three steps for $k=1,2,\ldots$ until convergence
\begin{align}
\label{App_Eq_opt_admm1}
\V^{(k)} &= \argmin{\{\V : \mathrm{diag}(\V)=\mathbf{1}_L \}} \,\,\, -\mathrm{Tr}( \hat{\Z}_n^H \hat{\Z}_n \V ) \nonumber \\ & ~~~~~~~~~~~~~~~~~~ + \rho \|\V - \tilde{\V}^{(k-1)} + \Y^{(k-1)} \|_F^2, \\
\label{App_Eq_opt_admm2}
\tilde{\V}^{(k)} &= \argmin{\{ \tilde{\V} : \tilde{\V} \succcurlyeq 0 \}} \,\,\, \| \V^{(k)} + \Y^{(k-1)} - \tilde{\V} \|_F^2,  \\
\label{App_Eq_opt_admm3}
\Y^{(k)} &= \Y^{(k-1)} + \V^{(k)} - \tilde{\V}^{(k)}.
\end{align}}

\tomtr{
Note that the first sub-problem can be easily solved by the projected gradient descent. Namely, by initializing $\V_0 = \V^{(k-1)}$ and repeating the following iterations for $q=1,2,\ldots$ until convergence
\begin{align}
\label{App_Eq_opt_admm1_pgd}
\V_q' &= \V_{q-1} - \gamma  ( -\hat{\Z}_n^H \hat{\Z}_n + \rho ( \V_{q-1} - \tilde{\V}^{(k-1)} + \Y^{(k-1)} )  )  \nonumber \\
\V_q &= \mathcal{P}_{\{\V : \mathrm{diag}(\V)=\mathbf{1}_L \}} \left \{  \V_q'  \right \},
\end{align}
where $\mathcal{P}_{\{\V : \mathrm{diag}(\V)=\mathbf{1}_L \}} \left \{ \V \right \} = \I_L + (\mathbf{1}_L\mathbf{1}_L^T-\I_L) \odot \V$ is the required the projection ($\odot$ denotes the element-wise product) and $\gamma$ is a step-size. Setting $\gamma$ to $1/\rho$ (which is one over the Lipschitz constant of the gradient of the first sub-problem) ensures convergence \cite{beck2009fast}.}

\tomtr{
The second sub-problem describes the projection of $\V^{(k)} + \Y_n^{(k-1)}$ onto positive semidefinite cone $\{ \V : \V \succcurlyeq 0 \}$, which, as explained above, is obtained by eigen-decomposing the symmetric version of $\V^{(k)} + \Y^{(k-1)}$ and setting the negative eigenvalues to zero. Namely,
\begin{align}
\label{App_Eq_opt_admm2_solution}
\tilde{\V}^{(k)} &= \mathcal{P}_{\{ \tilde{\V} : \tilde{\V} \succcurlyeq 0 \}} \left \{  \V^{(k)} + \Y^{(k-1)}  \right \} \nonumber \\
&= \sum_{\ell=1}^L \mathrm{max}\{\lambda_{\ell}, 0\} \u_{\ell}\u_{\ell}^H,
\end{align}
where $\lambda_{\ell}$ and $\u_{\ell}$ denote the $\ell$th eigenvalue and eigenvector of $\frac{(\V^{(k)} + \Y^{(k-1)})+(\V^{(k)} + \Y^{(k-1)})^H}{2}$, respectively. 
Note that since we initialize the ADMM with zero matrices (which are symmetric) and the gradient update and projection in the first sub-problem are symmetric, the symmetrization $\V^{(k)} + \Y^{(k-1)}$ is not required in our case.}

\tomtr{
The high-level description of the optimization scheme, with additional implementation details, is presented in Algorithm~\ref{opt_alg_sync}. Note that by-design only one eigen-decomposition is required in each ADMM iteration. Also, empirically, the stopping criteria of the projected gradient descent is met after a couple of iterations. Thus, Algorithm~\ref{opt_alg_sync} is extremely fast. In our experiments, using Intel-i7 laptop, it solved \eqref{Eq_opt_prob_convex_alpha} in about 0.02s (compared to about 0.2s when using the CVX).}




\bibliographystyle{ieeetr}
\bibliography{main_before_Rev2_arxiv} 

\end{document}